\documentclass[twocolumn]{revtex4}
\usepackage{graphicx}
\usepackage{bm}
\usepackage{graphicx}
\usepackage{amsmath}
\usepackage{mathrsfs}
\usepackage{ulem}
\usepackage{color}
\usepackage[colorlinks, linkcolor=red,anchorcolor=green,citecolor=blue]{hyperref}
\usepackage{dcolumn,booktabs,bm}
\usepackage{amsfonts,amssymb,stmaryrd,latexsym,amsmath}
\usepackage{textcomp}
\usepackage{multirow}
\usepackage{diagbox}
\usepackage{array}

\graphicspath{{./figs/}}

\begin{document}

\title{Cold and hot nuclear matter effects on $J/\psi$ production at RHIC-BES energies}

\author{Jiaxing Zhao}
\affiliation{Physics Department, Tsinghua University, Beijing 100084, China}
\author{Pengfei Zhuang}
%\email{}
\affiliation{Physics Department, Tsinghua University, Beijing 100084, China}

\begin{abstract}
Integrated and differential $J/\psi\ R_{AA}$ are systematically studied in Au-Au collisions at RHIC-BES energies in a transport approach, including cold and hot nuclear effects respectively in the initial condition and the collision terms. With decreasing energy, while the temperature, life time and size of the QGP fireball decrease, the nuclear absorption of the initially produced charmonia is more and more strong, and the nuclear shadowing effect on charmonium regeneration goes to anti-shadowing first and then to shadowing again. As a competition between the cold and hot nuclear effects, the QGP phase is still important for charmonium production at $\sqrt{s_{NN}}$=200, 62.4, 54.4 and 39 GeV but becomes negligible at $\sqrt{s_{NN}}$=14.5 GeV.
\end{abstract}

\pacs{...}
\maketitle

%=====================
\section{Introduction}
\label{s1}
It is widely accepted that a new state of matter, the so-called quark-gluon plasma (QGP), can be created in relativistic nuclear collisions. When the colliding energy is extremely high, like the collisions at Large Hadron Collider (LHC) and Relativistic Heavy Ion Collider (RHIC), the high temperature properties of the QGP is systematically studied for decades~\cite{STAR:2005gfr,Muller:2012zq}. Among the signatures of hot QGP, $J/\psi$ has long been considered as a sensitive hard probe~\cite{Matsui:1986dk}. The competition between suppression due to color screening and regeneration due to quark coalescence can explain almost all the experimentally measured $J/\psi$ data at low and intermediate momenta, including the nuclear modification factor $R_{AA}$~\cite{Zhou:2014kka,Chen:2018kfo,Zhao:2010nk,Zhao:2011cv,Liu:2009wza,Du:2015wha,Zhao:2017yan} and collective flows $v_2$~\cite{Zhu:2004nw,Zhou:2014kka,Chen:2018kfo,He:2021zej} and $v_3$~\cite{Zhao:2021voa}. 

The high density behavior of the QGP is still an open question and the main task of the Beam Energy Scan program at RHIC (RHIC-BES)~\cite{Bzdak:2019pkr}. When the colliding energy goes down, the life time and size of the produced QGP and its temperature are all reduced, but on the other hand, the cold nuclear matter effect on charmonium production becomes significant. For instance, the nuclear absorption is enhanced because of the longer collision time~\cite{Brambilla:2010cs}, and the nuclear effect on parton distribution goes from shadowing to anti-shadowing with decreasing energy in the RHIC-BES region~\cite{Helenius:2012wd}. When the colliding energy is low enough, the hot medium may disappear and the cold nuclear effect will dominant charmonium production, like the case in proton-nucleus collisions~\cite{Liu:2013via,Chen:2016dke,Du:2018wsj}. Considering the high statistics at RHIC-BES~\cite{STAR:2016utm} and the canonical enhancement effect on charmonium production~\cite{Liu:2013via,Gorenstein:2000ck}, one may precisely distinguish between the cold and hot nuclear matter effects. In this paper we focus on the cold and hot medium effects on $J/\psi$ production in the RHIC-BES energy region in a transport approach. 

%=====================
\section{$J/\psi$ transport in QGP}
\label{s2}
The charmonium motion in phase space can be described by a transport equation including both initial production via hard processes and regeneration in hot medium. The charmonium distribution $f_\psi({\bm p},x)$ for $\psi=J/\psi,\ \chi_c,\ \psi'$ in a phase-space cell with momentum ${\bm p}$ and space-time coordinate $x=(t,{\bm x})$ is controlled by a relativistic Boltzmann transport equation,
\begin{eqnarray}
\label{boltzmann}
p^\mu \partial_\mu f_\psi = -C_{loss} f_\psi + C_{gain}.
\end{eqnarray}
The hot nuclear effect, namely the charmonium suppression and regeneration in the created hot medium, is reflected in the loss term $C_{loss}$ and gain term $C_{gain}$. Considering that the medium effect in QGP phase is much stronger than that in the followed hadron phase even at lower colliding energies~\cite{Chen:2015ona}, we do not take the hadron phase into account in this work. In this case the anomalous suppression comes from the charmonium melting in the hot QCD medium via Debye screening~\cite{Digal:2005ht} and scattering with the surrounding partons~\cite{Peskin:1979va,Bhanot:1979vb,Grandchamp:2001pf}. Due to the Debye screening, the interaction range between a pair of charm quarks becomes shorter and shorter, and the bound state $(c\bar c)$ will disappear sequentially~\cite{Satz:2005hx}. This gives the dissociation temperatures $T_d\simeq(2.3,\ 1.2,\ 1.1)T_c$ for $J/\psi$, $\chi_c$, and $\psi'$. The dynamic scattering includes gluon dissociation and inelastic scatterings. In this paper, we consider the gluon dissociation $g+\psi\to c+\bar c$ as the dominant dissociation process in the QGP. The cross-section $\sigma_{g\psi}^{c\bar c}$ in vacuum can be derived through the operator production expansion method and was calculated firstly by Peskin and Bhanot~\cite{Peskin:1979va,Bhanot:1979vb}. Considering the inverse process $c+\bar c\to\psi+g$ for the gain term, the dissociation and regeneration rates $\alpha = C_{loss}/E$ and $\beta = C_{gain}/E$ are related to each other via detailed balance principal and can be expressed as~\cite{Zhao:2020jqu}
\begin{eqnarray}
\label{alphabeta}
\alpha({\bm p},x) &=& {1\over 2E}\int{d^3{\bm p}_g\over(2\pi)^32E_g}W_{g\psi}^{c\bar c}(s)f_g({\bm p}_g,x)\nonumber\\
&\times& \Theta(T(x)-T_c),\nonumber\\
\beta({\bm p},x) &=& {1\over 2E}\int {d^3{\bm p}_g\over(2\pi)^32E_g}{d^3{\bm p}_c \over(2\pi)^32E_c}{d^3{\bm p}_{\bar c}\over(2\pi)^32E_{\bar c}} \nonumber\\
&\times& F_c F_r W_{c\bar c}^{g\psi}(s) f_c({\bm p}_c,x)f_{\bar c}({\bm p}_{\bar c},x)\nonumber\\
&\times& \Theta(T(x)-T_c)(2\pi)^4\delta(p+p_g-p_c-p_{\bar c}),
\end{eqnarray}
where $W_{g\psi}^{c\bar c}(s)$ is the dissociation probability as a function of the center-of-mass energy $s=(p+p_g)^2$, constructed by the cross section $\sigma_{g\psi}^{c\bar c}(s)$ and related to the regeneration probability $W_{c\bar c}^{g\psi}(s)$ via detail balance, and $p=(E,{\bm p})\ p_g=(E_g,{\bm p}_g)\ p_c=(E_c,{\bm p}_c), p_{\bar c}=(E_{\bar c},{\bm p}_{\bar c})$ are charmonium, gluon, charm quark and anti-charm quark four-momentum. The temperature and baryon chemical potential dependence of the cross sections is reflected in the charmonium binding energy $\epsilon_\psi(T,\mu_B)$ which can be solved from the two-body Schr\"odinger equation in medium~\cite{Zhao:2020jqu}. In our numerical calculation we simply take $\epsilon_{J/\psi}= 150$ MeV~\cite{Liu:2009wza} which is smaller than its vacuum value. The step function $\Theta(T(x)-T_c)$ means that we consider here only the charmonium suppression and regeneration in the QGP phase, where $T_c$ is the deconfinement phase transition temperature determined by the equation of state of the system. The space-time evolution of the medium temperature $T(x)$ is solved from the hydrodynamics which will be discussed in the next section. 

Since gluons are a kind of constituents of the QGP, the gluon distribution is taken as the Bose-Einstein function $f_g=1/(e^{p\cdot u/T}-1)$ with local temperature $T(x)$ and velocity $u_\mu(x)$ of the medium controlled by the hydrodynamics. Considering the energy loss during the motion, the charm quark distribution should be controlled by a transport approach and in between two limits: the perturbative QCD limit without interaction with the medium and the thermalization limit via strong interaction with the medium. From the experimentally observed large charmed meson flow in Au-Au collisions at $\sqrt{s_{NN}}=200$ GeV~\cite{STAR:2017kkh}, charm quarks seem thermalized. We take, as a first approximation, a kinetically equilibrated distribution for charm (anti-charm) quarks $f_c=\rho_c(x)N(x)/(e^{p\cdot u/T}+1)$ with the local normalization factor $N(x)$. Considering the lower temperature and shorter QGP lifetime in nuclear collisions at RHIC-BES energies in comparison with the collisions at LHC or top RHIC energies, charm quarks may not fully thermalized. The partial thermalization can be described by introducing a relaxation factor~\cite{Song:2012at,Grandchamp:2002wp,Grandchamp:2003uw}
\begin{equation}
\label{fr}
F_r=1-e^{-\tau/\tau_r},
\end{equation}
where $\tau=\sqrt{t^2-z^2}$ is the proper time, which together with the longitudinal space-time rapidity $\eta=1/2\ln[(t-z)/(t+z)]$ are usually used to replace $t$ and $z$. The relaxation time $\tau_r$ characterizes the averaged thermalization time of the medium, and we take $\tau_r\simeq$ 7 fm/c~\cite{Zhao:2007hh} at RHIC-BES energies. The other effect on charm quark distribution at high baryon density is the canonical enhancement. When only few pairs of charm quarks are produced in an event, the charm conservation effect within the canonical ensemble needs to be considered. It becomes significant and enhances charmonium production in heavy ion collisions at lower energies. If the charm quark pairs are produced at the same rapidity, the canonical enhancement factor for charmonium production can be simply parameterized as~\cite{Liu:2013via,Gorenstein:2000ck,Andronic:2006ky,Kostyuk:2005zd} 
\begin{equation}
\label{fc}
F_c = 1+{1\over dN_{c\bar c}/dy}
\end{equation}
controlled by the number of directly produced $c\bar c$ pairs. We take $dN_{c\bar c}/dy = $1.31 and 0.027 in Au-Au collisions with centrality bin 0-60\% at $\sqrt{s_{NN}}$=200 and 14.5\ GeV, the corresponding canonical factor is 1.76 and 38.4. The charmonium enhancement is really dramatic in lower energy nuclear collisions. The charm quark density $\rho_c(x)$ in coordinate space is controlled by the charm conservation equation $\partial_\mu(\rho_c u^\mu)=0$. Since charm quarks at RHIC-BES energies are all produced via initial binary collisions and move freely in the pre-hydro stage, the initial density is governed by the nuclear geometry,
\begin{equation}
\label{rhoc}
\rho_c({\bm x},\tau_0) = {\mathcal T_A({\bm x}_T+{{\bm b}\over 2})\mathcal T_B({\bm x}_T-{{\bm b}\over 2})\cosh \eta \over \tau_0}{d\sigma_{pp}^{c\bar c}\over d\eta},
\end{equation}
where $\mathcal T_A$ and $\mathcal T_B$ are the thickness functions for the two colliding nuclei A and B~\cite{Miller:2007ri}, 
\begin{eqnarray}
\label{thickness}
\mathcal T_{A/B}({\bm x}_T) &=& T_{A/B}({\bm x}_T,-\infty,+\infty),\nonumber\\
T_{A/B}({\bm x}_T,z_i,z_j) &=& \int_{z_i}^{z_j}\rho_{A/B}({\bm x}_T, z)dz
\end{eqnarray}
with $\rho_{A/B}$ being the nuclear matter distribution in the two colliding nuclei, ${\bm x}_T$ is the transverse coordinate, ${\bm b}$ is the impact parameter of the nuclear collisions, and $d\sigma^{c\bar c}_{pp}/d\eta$ is the rapidity distribution of charm quark production cross section in p-p collisions~\cite{ALICE:2011aa,Cacciari:1998it,Cacciari:2001td}. The value of $d\sigma^{c\bar c}_{pp}/d\eta$ at RHIC-BES energies is shown in Table \ref{table1}.
%--------------------------------------------------------------
\begin{figure}[htb]
	{ $$\includegraphics[width=0.4\textwidth] {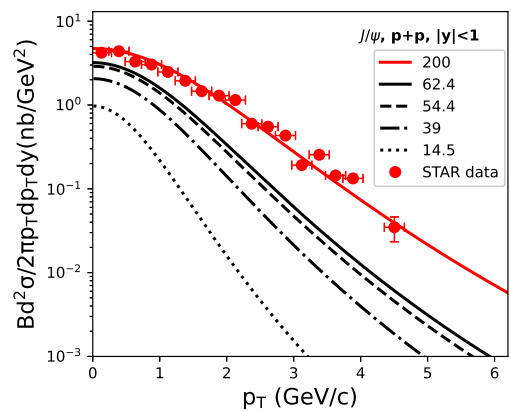}$$
		\caption{The $J/\psi\ p_T$ distribution at mid-rapidity in p-p collisions at RHIC-BES energies. The experiment data at $\sqrt{s_{NN}}=$ 200 GeV are from Ref.\cite{STAR:2018smh}, and $B(J/\psi\to e^+e^-)=5.97\pm0.03\%$~\cite{ParticleDataGroup:2020ssz} is the branch ratio.}
		\label{fig1}}
\end{figure}
%--------------------------end--------------------------------
%---------------------------------------------------------------------
\begin{table*}
	\renewcommand\arraystretch{1.5}
	\caption{The values of the used parameters to describe the initial medium ($\tau_0,\ s_0,\ \sigma_{pp}^{in}$), initial charmonium distribution ($d\sigma_{pp}^{J/\psi}/dy,\ \langle p_T^2\rangle,\ d\sigma_{pp}^{c\bar c}/dy$) and cold nuclear matter effect ($\sigma_{abs}^{J/\psi},\ a_{gN}$) at RHIC-BES energies. \label{table1}}
	\setlength{\tabcolsep}{1.0mm}
	\begin{tabular}{c|ccc|cc|c|cc}
		\toprule[1pt]\toprule[1pt] 
		$\sqrt{s_{NN}}$(GeV)  &$\tau_0$(fm/c) & $s_0$ & $\sigma_{pp}^{in}$(mb) & $d\sigma^{J/\psi}_{pp}/dy$(nb) &  $\langle p_T^2\rangle$(GeV)$^2$ & $d\sigma^{c\bar c}_{pp}/dy$($\mu$b) & $\sigma_{abs}^{J/\psi}$(mb)& $a_{gN}$(GeV$^2$/fm) \tabularnewline
		\midrule[1pt]
		200 & 0.6 & 9.3  & 41 & 716.7 & 3.05 & 162.0 & 1.5 & 0.100
		\tabularnewline
		62.4 & 1.0 & 10.8   & 36 & 295.7 & 1.85 & 45.0 & 4.8 & 0.085
		\tabularnewline
		54.4 & 1.1 & 10.5 & 35 & 252.0 & 1.74 & 38.3 & 5.0 & 0.085
		\tabularnewline
		39 & 1.3 &  10.35 & 34 & 150.7 & 1.46 & 27.5 & 5.2 & 0.080
		\tabularnewline
		14.5 & 2.2 & 9.22 & 32 & 37.6 & 0.77 & 3.3 & 8.9 & 0.077
		\tabularnewline
		\bottomrule[1pt]\bottomrule[1pt]
	\end{tabular}
\end{table*}
%---------------------------------------------------------------------

The initial charmonium distribution for the transport equation (\ref{boltzmann}) can be obtained from a superposition of p-p collisions, along with the modifications from several cold nuclear effects.
The $J/\psi$ momentum distribution in p-p collisions can be factorized as~\cite{Zha:2015eca},
\begin{equation}
\label{initial}
{d^2\sigma^{J/\psi}_{pp}\over 2\pi p_Tdp_Tdy}= {a\over 2\pi\langle p_T^2\rangle}\left(1+b^2{p_T^2\over \langle p_T^2\rangle}\right)^{-n} {d\sigma^{J/\psi}_{pp}\over dy}
\end{equation}
with parameters $a=2b^2(n-1),\ b=\Gamma(3/2)\Gamma(n-3/2)/\Gamma(n-1)$ and $n=3.93\pm0.03$. Taking the averaged momentum square and differential cross section shown in Table \ref{table1}, the $J/\psi$ transverse momentum distribution in p-p collisions at mid-rapidity is plotted in Fig.~\ref{fig1} at RHIC-BES energies and compared with the experimental data at $\sqrt{s_{\text{NN}}}=$ 200 GeV.

The cold nuclear effect usually consists of nuclear shadowing~\cite{Mueller:1985wy}, Cronin effect~\cite{Cronin:1974zm,Hufner:1988wz} and nuclear absorption~\cite{Gerschel:1988wn}. Unlike the collisions at LHC energy and top RHIC energy, the nuclear collision time at RHIC-BES energies is comparable with or even longer than the charmonium formation time, the produced charmonia will be sizeably absorbed by the surrounding nuclear matter. The surviving probability after the absorption can be expressed as
 \begin{equation}
 \label{absorption}
S_{abs}=e^{-\sigma_{abs}[T_A({\bm x}_T+{{\bm b}\over 2},z_A,\infty)+T_B({\bm x}_T-{{\bm b}\over 2},-\infty, z_B)]},
 \end{equation}
where $z_A$ and $z_B$ are the longitudinal coordinates of the charmonium production point in the local rest frames of the two colliding nuclei. The absorption cross section $\sigma_{abs}^{J/\psi}$ for the ground state $J/\psi$ is fixed by fitting the experimental data~\cite{Brambilla:2010cs}, see Table \ref{table1}. For the excited states, the experimental data are still rare. Neglecting the difference in formation time, the absorption cross sections for the excited states can be obtained from $\sigma_{abs}^{J/\psi}$ through the  mean-square-radius scaling law,
 \begin{equation}
\sigma_{abs}^\psi = {\langle r^2_\psi \rangle \over \langle r^2_{J/\psi} \rangle}\sigma_{abs}^{J/\psi},
 \end{equation}
where the mean-square-radius can be obtained by solving the two-body Schr\"odinger equation with Cornell potential~\cite{Zhao:2020jqu}, which leads to $\langle r^2_{J/\psi} \rangle=0.239~\text{fm}^2$, $\langle r^2_{\chi_c} \rangle=0.510~\text{fm}^2$, and $\langle r^2_{\psi(2S)} \rangle=0.808~\text{fm}^2$.

Before two gluons fuse into a charmonium, they acquire additional transverse momentum via multiscattering with the surrounding nucleons, and this extra momentum would be inherited by the produced charmonium, this is called Cronin effect~\cite{Cronin:1974zm,Hufner:1988wz}. Therefore, when doing the superposition of p-p distribution (\ref{initial}), we should make the replacement of $\langle p_T^2\rangle$ by~\cite{Huefner:2002tt} 
 \begin{equation}
 \label{cronin}
\langle p_T^2\rangle+a_{gN}l,
 \end{equation}
where the Cronin parameter $a_{gN}$ is the averaged charmonium transverse momentum square obtained from the gluon scattering with a unit of length of nucleons, and $l$ is the mean trajectory length of the two gluons in the two nuclei before the $c\bar c$ formation. The experimentally measured averaged momentum square for $J/\psi$ in p-A collisions at SPS energy and d-Au collisions at top RHIC energy can be well described by the Cronin effect~\cite{NA50:2003tdy,PHENIX:2007tnc}. The values of $a_{gN}$ at RHIC-BES energies are shown in Table \ref{table1}. To smooth the Cronin effect, we take in numerical calculations a Gaussian smearing~\cite{Huefner:2002tt,Zhao:2010nk} for the modified transverse momentum distribution.
%--------------------------------------------------------------
\begin{figure}[htb]
	{ $$\includegraphics[width=0.4\textwidth] {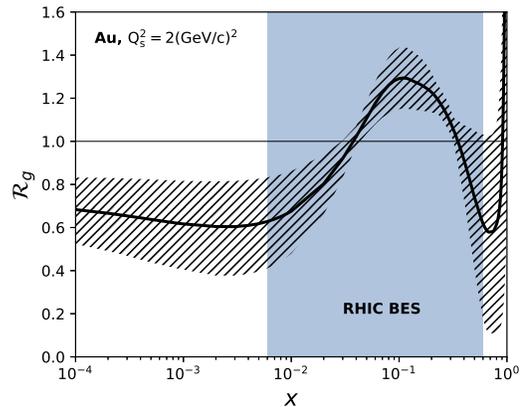}$$
		\caption{The nuclear modification factor $\mathcal {R}_g$ for gluons in nucleus Au from EPS09 NLO model~\cite{Helenius:2012wd}. The solid line inside the slashed band indicates the central value of the shadowing effect, and the vertical band shows the momentum fraction range covered by RHIC-BES energies for two gluons to produce a $J/\psi$.}
		\label{fig2}}
\end{figure}
%--------------------------end--------------------------------

The shadowing effect modifies the parton distribution in a nucleus relative to that in a nucleon, which changes the open and hidden charm yields in nuclear collisions~\cite{Gavin:1996bx,Vogt:2010aa}. The effect is mainly due to the parton collectivity in a nucleus~\cite{Norton:2003cb} and can be parameterized as a modification factor $\mathcal{R}_i = \bar f_i(x,\mu_F)/(Af_i(x,\mu_F))$, where $\bar f_i$ and $f_i$ are the parton distribution functions ($i$=gluon, light, strange, and heavy quarks) in a nucleus and a free nucleon, $x$ is the longitudinal momentum fraction, and the factorization scale is taken as $\mu_F=\sqrt{m_\psi^2+p_T^2}$. The modification factor can be simulated by, for instance, the EPS09 package~\cite{Helenius:2012wd}, as shown in Fig.~\ref{fig2} for gluons. The interesting point here is the $x$ dependence of the shadowing effect at RHIC-BES energies. A simple estimation of the longitudinal momentum fraction $x$ of the two initial gluons in this energy region are in the range of $0.006<x<0.607$, see the vertical band in Fig.\ref{fig2}. Around and above the top RHIC energy, there is always shadowing effect with $\mathcal {R}_g<1$, while in the RHIC-BES region the shadowing approaches to anti-shadowing when the colliding energy decreases! This change of shadowing effect may significantly affect the charmonium production in nuclear collisions.      

Including the above discussed nuclear absorption, Cronin effect and shadowing effect, the initial charmonium distribution can be written as
 \begin{eqnarray}
 \label{initial2}
&& f_\psi({\bm p}, {\bm x}, \tau_0)\nonumber\\
&=& {(2\pi)^3 \over E \tau_0}\int dz_A dz_B \rho_A({\bm x}_T+{{\bm b}\over 2},z_A)\rho_B({\bm x}_T-{{\bm b}\over 2},z_B)\nonumber\\
&\times& S_{abs} \mathcal{R}_g(x_1,\mu_F, {\bm x}_T+{{\bm b}\over 2})\mathcal{R}_g(x_2,\mu_F, {\bm x}_T-{{\bm b}\over 2})\nonumber\\
&\times& f^{pp}_\psi({\bm p},{\bm x},z_A,z_B).
 \end{eqnarray}

%=====================
\section{Evolution of QGP}
\label{s3}
The quark matter created in high energy nuclear collisions is a very prefect fluid~\cite{Teaney:2003kp,Lacey:2006bc} and its space-time evolution can be simulated by hydrodynamics with the conservation equations of energy-momentum and net baryon density,
 \begin{eqnarray}
 \label{hydro}
\partial_\mu T^{\mu \nu} &=& 0, \nonumber\\
\partial_\mu J_B^\mu &=& 0.
\end{eqnarray}
The energy-momentum tensor $T^{\mu \nu}$ and net baryon current $J_B^\mu$ are expressed as
 \begin{eqnarray}
 \label{tuvb}
T^{\mu \nu} &=& \epsilon u^\mu u^v-(P+\Pi)\Delta^{\mu \nu}+\pi^{\mu \nu}, \nonumber\\
J_B^\mu &=& n_Bu^\mu + q^\mu,
\end{eqnarray}
where $\epsilon$ is the medium energy density, $P$ the pressure, $\Pi$ the bulk viscous pressure, $\Delta^{\mu \nu}= g^{\mu \nu}-u^\mu u^\nu$ the projection tensor, $\pi^{\mu \nu}$ the shear stress tensor, $n_B$ the net baryon density, and $q^\mu$ the baryon diffusion current. In the following calculation, we take the ratio of shear viscosity to entropy density as a constant $\eta/s=0.08$, and neglect the bulk viscosity and baryon diffusion. To close the hydrodynamic equations, we use NEOS-B as the equation of state at finite baryon chemical potential~\cite{Monnai:2019hkn,Monnai:2021kgu}.

 %--------------------------------------------------------------
\begin{figure}[htb]
{\includegraphics[width=0.4\textwidth] {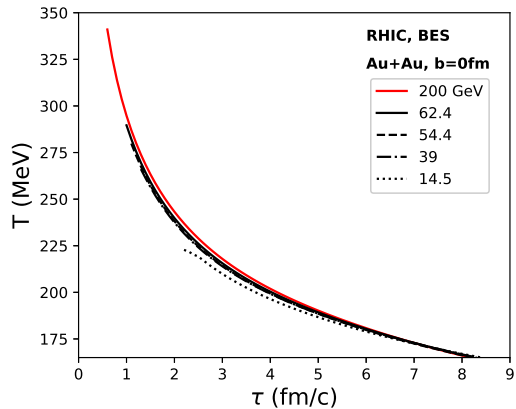}
\caption{The temperature evolution at the center of the QGP medium created in central Au-Au collisions at RHIC-BES energies. }
\label{fig3}}
\end{figure}
%--------------------------end--------------------------------
The initial condition of the hydrodynamic equations, such as the initial entropy density and baryon density, can be obtained from the assumption that they come independently from the two colliding nuclei~\cite{Denicol:2018wdp},
 \begin{eqnarray}
s({\bm x},\tau_0) &=& {s_0\over \tau_0}[f_P^s(\eta) N_P({\bm x}_T)+f_T^s(\eta) N_T({\bm x}_T)],\nonumber\\
n_B({\bm x},\tau_0) &=& {1\over \tau_0}[f_P^n(\eta) N_P({\bm x}_T)+f_T^n(\eta) N_T({\bf x}_T)],
\end{eqnarray}
where $s_0$ is the maximum entropy density which is adjusted to reproduce the experimentally observed multiplicity, and $f_P^s, f_P^n$ and $f_T^s, f_T^n$ are rapidity distributions of the initial entropy and baryon number produced by the projectile and target nuclei. Since the initial entropy mainly comes from the ``soft'' processes in low energy nuclear collisions~\cite{Kharzeev:2000ph}, the number of participating nucleons is the number of sources to produce entropy and baryon number $N_P$ and $N_T$ which can be calculated through the optical Glauber model~\cite{Denicol:2018wdp,Wu:2021fjf}, 
 \begin{eqnarray}
N_P({\bm x}_T) &=& \mathcal T_A({\bm x}_T+{{\bf b}\over 2})(1-e^{-\sigma_{pp}^{in}\mathcal T_B({\bm x}_T-{{\bf b}\over 2})}),\nonumber\\
N_T({\bm x}_T) &=& \mathcal T_B({\bm x}_T-{{\bf b}\over 2})(1-e^{-\sigma_{pp}^{in}\mathcal T_A({\bm x}_T+{{\bf b}\over 2})}).
\end{eqnarray}
The values of the inelastic scattering cross section $\sigma_{pp}^{in}$ in p-p collisions which can be obtained through theoretical calculation and experimental measurement~\cite{ALICE:2012fjm,Sjostrand:2006za}, the initial time $\tau_0$ of the medium which can be estimated from the overlap time of the two colliding nuclei~\cite{Shen:2017bsr,Shen:2017fnn}, and the maximum entropy density $s_0$ at RHIC-BES energies are listed in Table \ref{table1}.

The hydrodynamic equations with the above initial condition can be numerically solved with the help of the MUSIC package~\cite{Schenke:2010nt,Denicol:2018wdp}. The time evolution of the temperature at the center of the QGP medium created in central Au-Au collisions at RHIC-BES energies is shown in Fig.\ref{fig3}. Considering the expansion of the fluid, the temperature decreases monotonously with time. While the thermalization time (initial time) of the system is different at different colliding energies, the evolution trajectory of the temperature is almost the same for all the energies. It starts at the maximum temperature at initial time $\tau_0$ and ends at the critical temperature $T_c$ of the QGP. The maximum temperature for colliding energies $\sqrt{s_{NN}}$=200, 62.4, 54.4, 39, and 14.5 GeV is 321.0, 289.6, 279.8, 266.1, and 244.5 MeV, and the critical temperature $T_c=165$ MeV is almost colliding energy independent~\cite{Andronic:2008gu,Cleymans:2005xv}. Since the QGP phase is formed even at $\sqrt{s_{\text{NN}}}=14.5$ GeV, we can use the transport equation (\ref{boltzmann}) to describe charmonium motion in nuclear collisions at RHIC-BES energies.  

%=====================
\section{Results}
\label{s4}
The Boltzmann equation (\ref{boltzmann}), with cold nuclear effect in the initial condition (\ref{initial2}) and hot nuclear effect in the collision terms (\ref{alphabeta}), can be analytically solved~\cite{Liu:2009wza,Zhao:2020jqu}. As a ground state of $c\bar c$, the experimentally observed $J/\psi$s contain the direct production and the decay contribution from the excited states and B-hadrons. Since the B-dacay is mainly in high $p_T$ region and becomes important at LHC energy, we neglect it for nuclear collisions at RHIC-BES energies. For the feed-down, we take $22\%$ of $\chi_c$s and $61\%$ of $\psi'$s to decay to $J/\psi$s in this energy region~\cite{ParticleDataGroup:2020ssz}. We focus in this section on the $J/\psi$ nuclear modification factor $R_{AA} = N_{AA}/(N_{\text{coll}}N_{pp})$, where $N_{pp}$ and $N_{AA}$ are the $J/\psi$ numbers produced in p-p and Au-Au collisions, and $N_{part}$ and $N_{coll}$ are the number of participating nucleons and the number of binary collisions. All the calculations are at mid-rapidity.  
%--------------------------------------------------------------
\begin{figure}[htb]
	{ \includegraphics[width=0.4\textwidth] {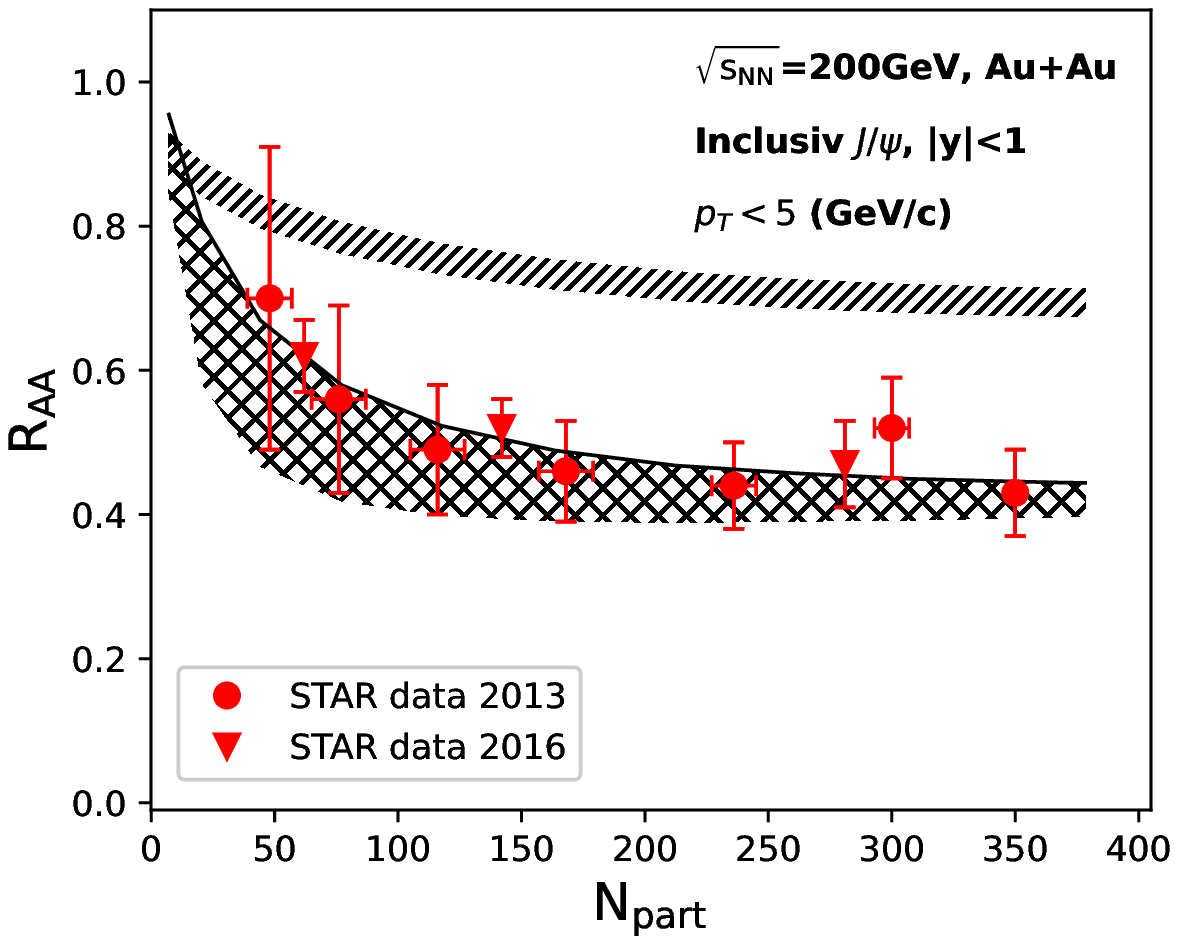}\\
		\includegraphics[width=0.4\textwidth] {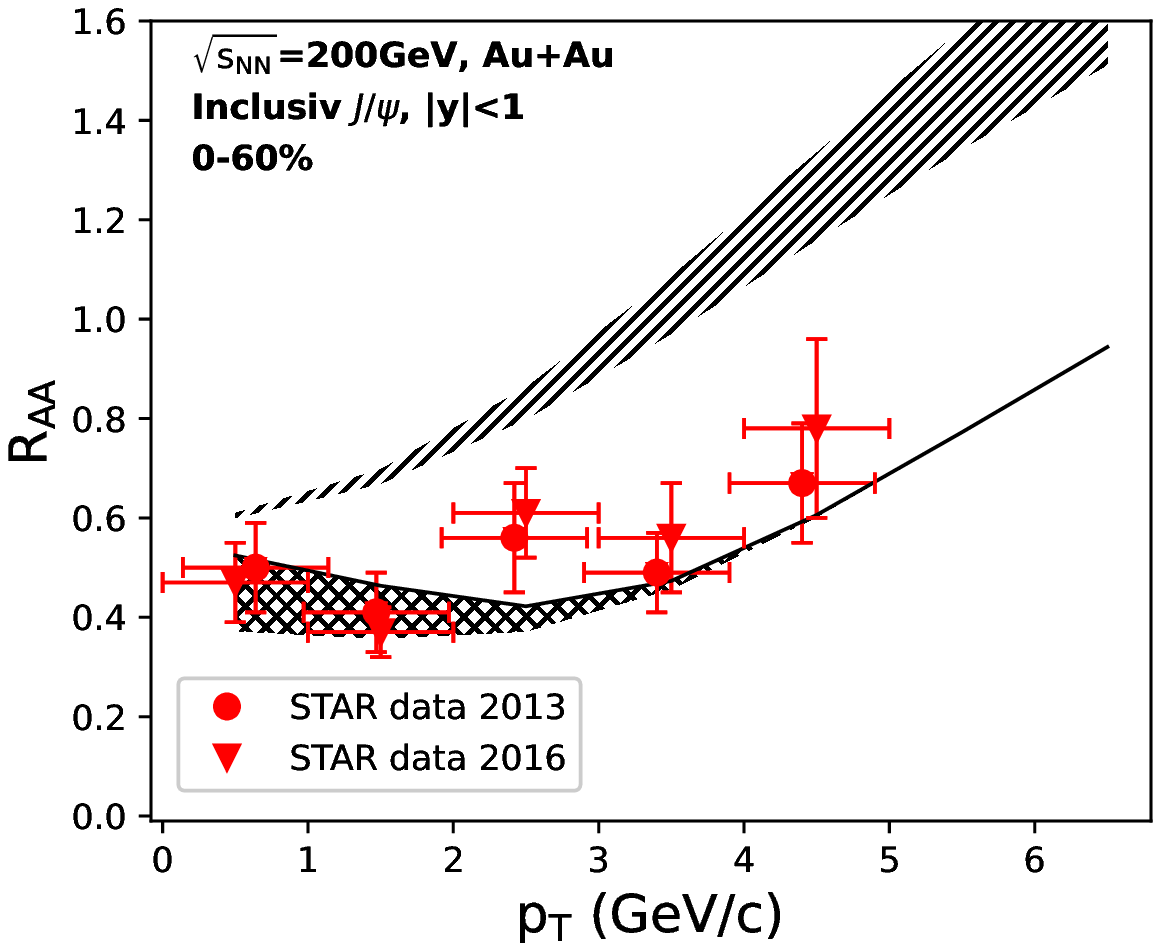}
		\caption{The nuclear modifications factor $R_{AA}$ for inclusive $J/\psi$s as a function of $N_{part}$ (upper panel) and $p_T$ (lower panel) in Au-Au collisions at top RHIC energy $\sqrt{s_{NN}}$=200 GeV. The slashed and crossed bands are the calculations with only cold nuclear effect and with both cold and hot nuclear effects. The experimental data are from the STAR collaboration~\cite{STAR:2016utm,STAR:2012wnc}.}
		\label{fig4}}
\end{figure}
%--------------------------end--------------------------------
%--------------------------------------------------------------
\begin{figure}[htb]
{ $$\includegraphics[width=0.48\textwidth] {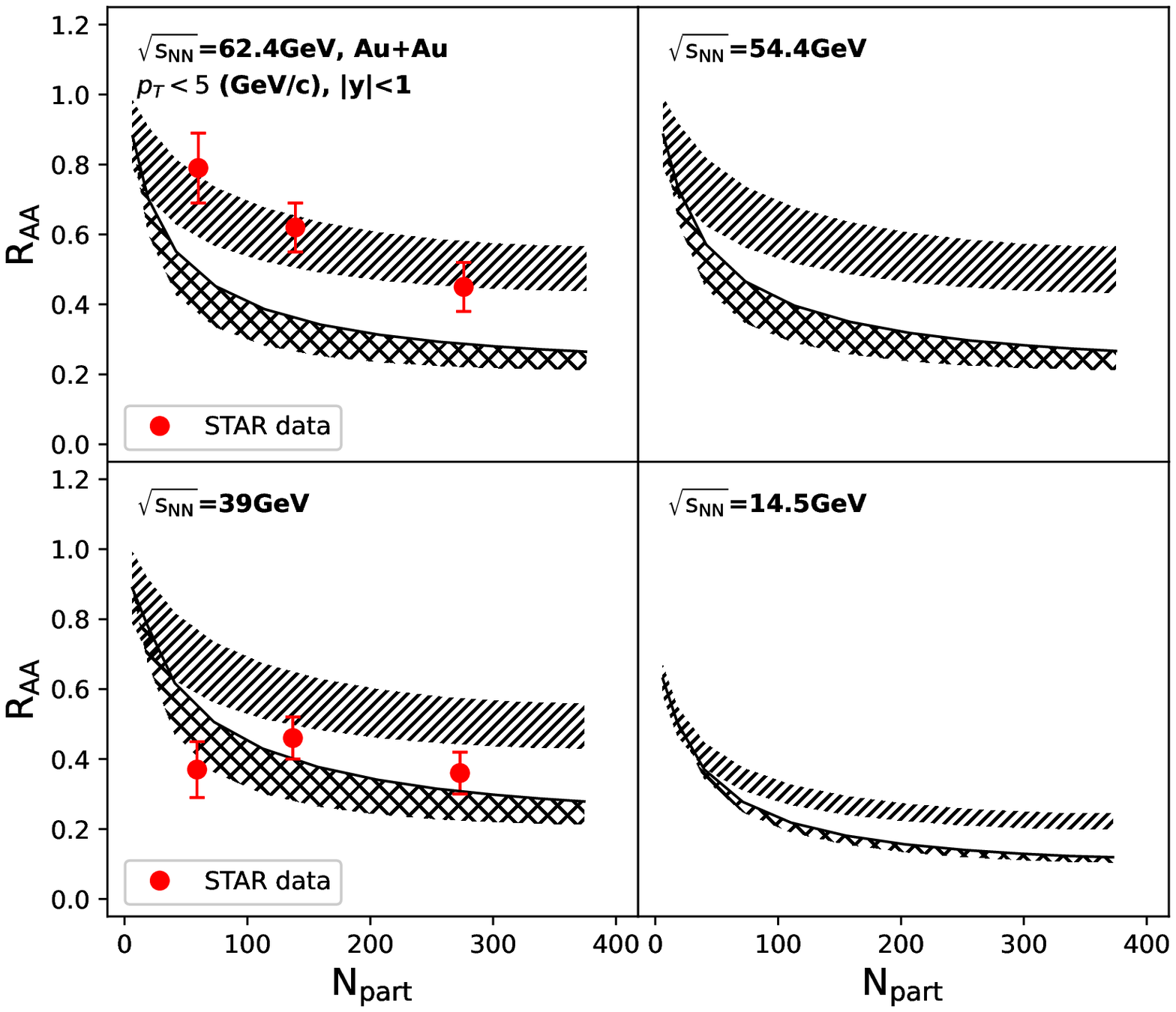}$$
\caption{The nuclear modifications factor $R_{AA}$ for inclusive $J/\psi$s as a function of $N_{part}$ in Au-Au collisions at colliding energies $\sqrt{s_{NN}}$=62.4, 54.4, 39 and 14.5 GeV. The slashed and crossed bands are the calculations with only cold nuclear effect and with both cold and hot nuclear effects. The experimental data are from the STAR collaboration~\cite{STAR:2016utm}.}
\label{fig5}}
\end{figure}
%--------------------------end--------------------------------
 %--------------------------------------------------------------
\begin{figure}[htb]
{ $$\includegraphics[width=0.48\textwidth] {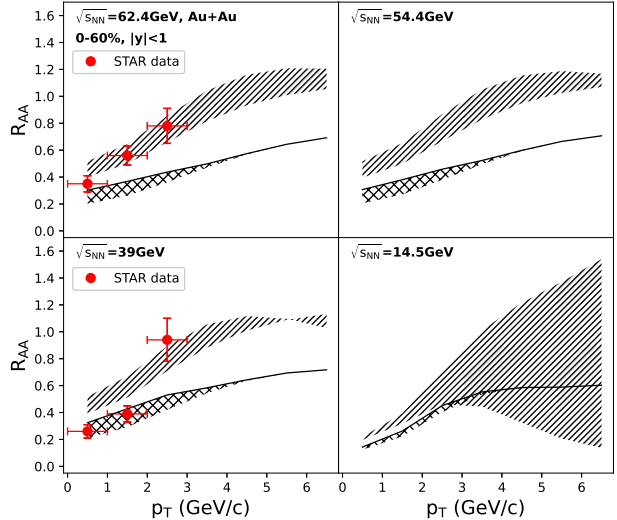}$$
\caption{The nuclear modifications factor $R_{AA}$ for inclusive $J/\psi$s as a function of $p_T$ in Au-Au collisions at colliding energies $\sqrt{s_{NN}}$=62.4, 54.4, 39 and 14.5 GeV. The slashed and crossed bands are the calculations with only cold nuclear effect and with both cold and hot nuclear effects. The experimental data are from the STAR collaboration~\cite{STAR:2016utm}.}
\label{fig6}}
\end{figure}
%--------------------------end--------------------------------

The centrality dependence of $R_{AA}$ and its transverse momentum distribution in a fixed centrality bin $0-60\%$ at RHIC top energy $\sqrt{s_{NN}}=200$ GeV are shown in Fig.\ref{fig4}. The slashed bands are the calculations considering only cold nuclear effect ($\alpha=\beta=0$). The band structure is due to the uncertainty of the shadowing effect shown in Fig.\ref{fig2}. The full result with both cold and hot nuclear effects is plotted as the crossed bands, where the shadowing effect is taken as its central value indicated by the solid line in Fig.~\ref{fig2}. The upper and lower limits of the bands are the calculations with and without considering the canonical enhancement factor $F_c$. Since the canonical effect influences only the regeneration process in peripheral collisions, it becomes important at lower $N_{part}$ and lower $p_T$. At high $p_T$, the regeneration contribution and in turn the canonical enhancement disappears, the band approaches to a line. The competition between the two aspects of the hot nuclear effect, namely the suppression and regeneration, lows down the charmonium yield. From the comparison with the experimental data~\cite{STAR:2016utm,STAR:2012wnc}, the QGP effect which is the difference between the slashed and crossed bands, is significant at RHIC top energy. 
 %--------------------------------------------------------------
\begin{figure}[htb]
{ $$\includegraphics[width=0.4\textwidth] {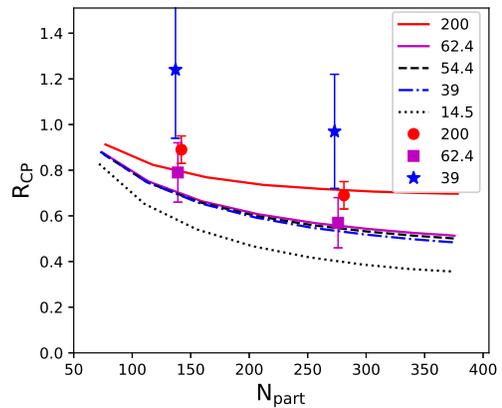}$$
\caption{The nuclear modification factor $R_{CP}$ for inclusive $J/\psi$s as a function of centrality in Au-Au collisions at RHIC-BES energies. The peripheral collision is defined as the collisions with centrality bin 40-60\%, and the experimental data are from the STAR Collaboration~\cite{STAR:2016utm}.}
\label{fig7}}
\end{figure}
%--------------------------end--------------------------------

We now turn to see the energy dependence of charmonium production. The $J/\psi$ nuclear modification factor $R_{AA}$ as a function of centrality $N_{part}$ and transverse momentum $p_T$ are displayed in Figs.\ref{fig5} and \ref{fig6} at colliding energies $\sqrt{s_{NN}}$= 62.4, 54.4, 39, and 14.5 GeV. Again the slashed and crossed bands are the calculations with only cold nuclear effect and with both cold and hot nuclear effects. When the colliding energy decreases from 62.4 to 54.4 and then to 39 GeV, the anti-shadowing effect becomes more and more strong, see Fig.\ref{fig2}, which enhances the charmonium production and compensates for the larger and larger nuclear absorption. As a result, the cold nuclear effect is almost the same at these three energies. When the energy decreases further, the anti-shadowing comes back to shadowing, and the averaged modification factor $\mathcal R_g$ at 14.5 GeV is slightly less than unit, and the strong nuclear absorption in this case leads to a strong $J/\psi$ suppression. While the canonical enhancement factor $F_c$ is extremely large at $\sqrt{s_{NN}}$=14.5 GeV, its contribution to the total yield is very small, see the narrow bands in Figs.\ref{fig5} and \ref{fig6}. This is due to the fact that, the canonical effect modifies only the regeneration and the regeneration part becomes less and less important with decreasing colliding energy. At $\sqrt{s_{NN}}=$ 62.4, 54.4 and 39 GeV, the hot nuclear effect, namely the difference between the slashed and corresponding crossed band, is still very clear, which indicates a sizeable QGP formation in the collisions at these energies. However, when the colliding energy goes down to 14.5 GeV, the hot nuclear effect is already weak, the two bands in $p_T$ distribution becomes even indistinguishable due to the large uncertainty in the shadowing effect. From the comparison with the experimental data, the full calculation agrees reasonably well with the data at $\sqrt{s_{NN}}$=200 and 39 GeV, but it deviates clearly from the data at $\sqrt{s_{NN}}$=62.4 GeV, for both $N_{part}$ and $p_T$ distributions. It is strange that only the cold nuclear effect can explain the data very well at this intermediate energy, while the hot nuclear effect plays an important role at higher and lower energies. The case here is similar to other model calculations~\cite{Zhao:2010nk,STAR:2016utm}. More precise theoretical and experimental study around this energy is needed.  

Aiming to eliminate the uncertainty from p-p collisions, one usually defines the nuclear modification factor $R_{CP}$ which is a ratio of the particle yield in collisions with a given centrality bin to that in the peripheral collisions, $R_{CP} = (dN/dy/N_{coll})|_b/(dN/dy/N_{coll})_{perp}$. For $J/\psi$, this is a quantity to describe the relative suppression in the two bins. If the medium effect is centrality independent, there is always $R_{CP}=1$. With the above described cold and hot medium effects, $R_{CP}$ as a function of centrality and the comparison with experimental data are shown in Fig.\ref{fig7} for Au-Au collisions at RHIC-BES energies. The deviation from unit indicates clearly the sizeable medium effect at these energies.  
 
%=====================
\section{Summary}\label{summary}
Cold and hot nuclear matter effects are the driving force to study nuclear collisions at high energies. The former is the basement and the latter is the condition to form the QGP, a new state of nuclear matter. At LHC and top RHIC energies, the cold nuclear effect is weak and the hot nuclear effect is the dominant one. For heavy ion collisions at RHIC-BEC energies, the cold nuclear effect becomes strong and the hot nuclear effect is still important, the competition between the two governs the charmonium production as a probe of the QGP. In this paper we take a transport equation, which can distinguish clearly the cold nuclear effect in the initial condition from the hot nuclear effect in the collision terms, to study integrated and differential $J/\psi$ $R_{AA}$ in Au-Au collisions at RHIC-BES energies. 

With decreasing energy, more and more initially produced charmonia are absorbed by the surrounding nuclear matter, and the shadowing effect on charmonium regeneration becomes anti-shadowing first and then shadowing again. This change in nuclear absorption and shadowing leads to an almost energy-independent cold nuclear effect at $\sqrt{s_{NN}}=$62.4, 54.4 and 39 GeV and a much stronger cold nuclear effect at $\sqrt{s_{NN}}$=14.5 GeV. For the hot nuclear effect, the temperature, life time and size of the QGP fireball monotonously decrease with decreasing colliding energy. At $\sqrt{s_{NN}}$=14.5 GeV the cold nuclear effect becomes the dominant one and the hot nuclear effect is negligible.  \\

\noindent {\bf Acknowledgement}: The work is supported by Guangdong Major Project of Basic and Applied Basic Research No. 2020B0301030008 and NSFC grant Nos. 11890712, 12075129, and 12175165.  
%=====================


\begin{thebibliography}{99}

%\cite{STAR:2005gfr}
\bibitem{STAR:2005gfr}
J.~Adams \textit{et al.} [STAR],
%``Experimental and theoretical challenges in the search for the quark gluon plasma: The STAR Collaboration's critical assessment of the evidence from RHIC collisions,''
Nucl. Phys. A \textbf{757}, 102-183 (2005).
%doi:10.1016/j.nuclphysa.2005.03.085
%[arXiv:nucl-ex/0501009 [nucl-ex]].
%3428 citations counted in INSPIRE as of 03 Jan 2022

%\cite{Muller:2012zq}
\bibitem{Muller:2012zq}
B.~Muller, J.~Schukraft and B.~Wyslouch,
%``First Results from Pb+Pb collisions at the LHC,''
Ann. Rev. Nucl. Part. Sci. \textbf{62}, 361-386 (2012).
%doi:10.1146/annurev-nucl-102711-094910
%[arXiv:1202.3233 [hep-ex]].
%416 citations counted in INSPIRE as of 03 Jan 2022

%\cite{Matsui:1986dk}
\bibitem{Matsui:1986dk}
T.~Matsui and H.~Satz,
%``$J/\psi$ Suppression by Quark-Gluon Plasma Formation,''
Phys. Lett. B \textbf{178}, 416-422 (1986).
%doi:10.1016/0370-2693(86)91404-8
%3208 citations counted in INSPIRE as of 04 Oct 2021

%\cite{Zhao:2011cv}
\bibitem{Zhao:2011cv}
X.~Zhao and R.~Rapp,
%``Medium Modifications and Production of Charmonia at LHC,''
Nucl. Phys. A \textbf{859}, 114-125 (2011).
%doi:10.1016/j.nuclphysa.2011.05.001
%[arXiv:1102.2194 [hep-ph]].
%236 citations counted in INSPIRE as of 13 Jun 2020

%\cite{Zhao:2010nk}
\bibitem{Zhao:2010nk}
X.~Zhao and R.~Rapp,
%``Charmonium in Medium: From Correlators to Experiment,''
Phys. Rev. C \textbf{82}, 064905 (2010).
%doi:10.1103/PhysRevC.82.064905
%[arXiv:1008.5328 [hep-ph]].
%158 citations counted in INSPIRE as of 04 Oct 2021

%\cite{Liu:2009wza}
\bibitem{Liu:2009wza}
Y.~Liu, Z.~Qu, N.~Xu and P.~Zhuang,
%``Rapidity Dependence of J/psi Production at RHIC and LHC,''
J. Phys. G \textbf{37}, 075110 (2010).
%doi:10.1088/0954-3899/37/7/075110
%[arXiv:0907.2723 [nucl-th]].
%23 citations counted in INSPIRE as of 13 Oct 2021

%\cite{Du:2015wha}
\bibitem{Du:2015wha}
X.~Du and R.~Rapp,
%``Sequential Regeneration of Charmonia in Heavy-Ion Collisions,''
Nucl. Phys. A \textbf{943}, 147-158 (2015).
%doi:10.1016/j.nuclphysa.2015.09.006
%[arXiv:1504.00670 [hep-ph]].
%107 citations counted in INSPIRE as of 13 Oct 2021

%\cite{Zhao:2017yan}
\bibitem{Zhao:2017yan}
J.~Zhao and B.~Chen,
%``Strong diffusion effect of charm quarks on J /$\psi$ production in Pb\textendash{}Pb collisions at the LHC,''
Phys. Lett. B \textbf{776}, 17-21 (2018).
%doi:10.1016/j.physletb.2017.11.014
%[arXiv:1705.04558 [nucl-th]].
%21 citations counted in INSPIRE as of 04 Jan 2022

%\cite{Zhou:2014kka}
\bibitem{Zhou:2014kka}
K.~Zhou, N.~Xu, Z.~Xu and P.~Zhuang,
%``Medium effects on charmonium production at ultrarelativistic energies available at the CERN Large Hadron Collider,''
Phys. Rev. C \textbf{89}, no.5, 054911 (2014).
%doi:10.1103/PhysRevC.89.054911
%[arXiv:1401.5845 [nucl-th]].
%112 citations counted in INSPIRE as of 11 Jun 2020

%\cite{Chen:2018kfo}
\bibitem{Chen:2018kfo}
B.~Chen,
%``Thermal production of charmonia in Pb-Pb collisions at $\sqrt {s_{NN}} = $ 5.02 TeV,''
Chin. Phys. C \textbf{43}, no.12, 124101 (2019).
%doi:10.1088/1674-1137/43/12/124101
%[arXiv:1811.11393 [nucl-th]].
%5 citations counted in INSPIRE as of 13 Oct 2021

%\cite{Zhu:2004nw}
\bibitem{Zhu:2004nw}
X.~l.~Zhu, P.~f.~Zhuang and N.~Xu,
%``J/psi transport in QGP and p(t) distribution at SPS and RHIC,''
Phys. Lett. B \textbf{607}, 107-114 (2005).
%doi:10.1016/j.physletb.2004.12.023
%[arXiv:nucl-th/0411093 [nucl-th]].
%76 citations counted in INSPIRE as of 21 Sep 2021

%\cite{He:2021zej}
\bibitem{He:2021zej}
M.~He, B.~Wu and R.~Rapp,
%``Collectivity of $J/\psi$ Mesons in Heavy-Ion Collisions,''
[arXiv:2111.13528 [nucl-th]].
%0 citations counted in INSPIRE as of 04 Jan 2022

%\cite{Zhao:2021voa}
\bibitem{Zhao:2021voa}
J.~Zhao, B.~Chen and P.~Zhuang,
%``Charmonium Triangular Flow in High Energy Nuclear Collisions,''
[arXiv:2112.00293 [hep-ph]].
%0 citations counted in INSPIRE as of 03 Jan 2022

%\cite{Bzdak:2019pkr}
\bibitem{Bzdak:2019pkr}
A.~Bzdak, S.~Esumi, V.~Koch, J.~Liao, M.~Stephanov and N.~Xu,
%``Mapping the Phases of Quantum Chromodynamics with Beam Energy Scan,''
Phys. Rept. \textbf{853}, 1-87 (2020).
%doi:10.1016/j.physrep.2020.01.005
%[arXiv:1906.00936 [nucl-th]].
%153 citations counted in INSPIRE as of 04 Oct 2021

%\cite{Brambilla:2010cs}
\bibitem{Brambilla:2010cs}
N.~Brambilla, S.~Eidelman, B.~K.~Heltsley, R.~Vogt, G.~T.~Bodwin, E.~Eichten, A.~D.~Frawley, A.~B.~Meyer, R.~E.~Mitchell and V.~Papadimitriou, \textit{et al.}
%``Heavy Quarkonium: Progress, Puzzles, and Opportunities,''
Eur. Phys. J. C \textbf{71}, 1534 (2011).
%doi:10.1140/epjc/s10052-010-1534-9
%[arXiv:1010.5827 [hep-ph]].
%1636 citations counted in INSPIRE as of 29 Oct 2021

%\cite{Helenius:2012wd}
\bibitem{Helenius:2012wd}
I.~Helenius, K.~J.~Eskola, H.~Honkanen and C.~A.~Salgado,
%``Impact-Parameter Dependent Nuclear Parton Distribution Functions: EPS09s and EKS98s and Their Applications in Nuclear Hard Processes,''
JHEP \textbf{07}, 073 (2012).
%doi:10.1007/JHEP07(2012)073
%[arXiv:1205.5359 [hep-ph]].
%149 citations counted in INSPIRE as of 16 Sep 2021

%\cite{Liu:2013via}
\bibitem{Liu:2013via}
Y.~Liu, C.~M.~Ko and T.~Song,
%``Hot medium effects on $J/\psi$ production in $p+Pb$ collisions at $\sqrt{s_{NN}}=5.02$ TeV,''
Phys. Lett. B \textbf{728}, 437-442 (2014).
%doi:10.1016/j.physletb.2013.12.016
%[arXiv:1309.5113 [nucl-th]].
%19 citations counted in INSPIRE as of 08 Oct 2021

%\cite{Chen:2016dke}
\bibitem{Chen:2016dke}
B.~Chen, T.~Guo, Y.~Liu and P.~Zhuang,
%``Cold and Hot Nuclear Matter Effects on Charmonium Production in p+Pb Collisions at LHC Energy,''
Phys. Lett. B \textbf{765}, 323-327 (2017).
%doi:10.1016/j.physletb.2016.12.021
%[arXiv:1607.07927 [nucl-th]].
%25 citations counted in INSPIRE as of 03 Jan 2022

%\cite{Du:2018wsj}
\bibitem{Du:2018wsj}
X.~Du and R.~Rapp,
%``In-Medium Charmonium Production in Proton-Nucleus Collisions,''
JHEP \textbf{03}, 015 (2019).
%doi:10.1007/JHEP03(2019)015
%[arXiv:1808.10014 [nucl-th]].
%38 citations counted in INSPIRE as of 03 Jan 2022


%\cite{STAR:2016utm}
\bibitem{STAR:2016utm}
L.~Adamczyk \textit{et al.} [STAR],
%``Energy dependence of $J/\psi$ production in Au+Au collisions at $\sqrt{s_{NN}} =$ 39, 62.4 and 200 GeV,''
Phys. Lett. B \textbf{771}, 13-20 (2017).
%doi:10.1016/j.physletb.2017.04.078
%[arXiv:1607.07517 [hep-ex]].
%26 citations counted in INSPIRE as of 04 Oct 2021

%\cite{Gorenstein:2000ck}
\bibitem{Gorenstein:2000ck}
M.~I.~Gorenstein, A.~P.~Kostyuk, H.~Stoecker and W.~Greiner,
%``Statistical coalescence model with exact charm conservation,''
Phys. Lett. B \textbf{509}, 277-282 (2001).
%doi:10.1016/S0370-2693(01)00516-0
%[arXiv:hep-ph/0010148 [hep-ph]].
%160 citations counted in INSPIRE as of 08 Oct 2021

%\cite{Chen:2015ona}
\bibitem{Chen:2015ona} 
  B.~Chen, P.~Zhuang and Z.~Xu,
  %``Effects of quark-gluon plasma and hadron gas on charmonium production at energies available at the CERN Super Proton Synchrotron and the Facility for Antiproton and Ion Research,''
  Phys.\ Rev.\ C {\bf 93}, no. 4, 044917 (2016). 
  
%\cite{Digal:2005ht}
\bibitem{Digal:2005ht}
S.~Digal, O.~Kaczmarek, F.~Karsch and H.~Satz,
%``Heavy quark interactions in finite temperature QCD,''
Eur. Phys. J. C \textbf{43}, 71-75 (2005).
%doi:10.1140/epjc/s2005-02309-7
%[arXiv:hep-ph/0505193 [hep-ph]].
%76 citations counted in INSPIRE as of 03 Jan 2022

%\cite{Grandchamp:2001pf}
\bibitem{Grandchamp:2001pf}
L.~Grandchamp and R.~Rapp,
%``Thermal versus direct J / Psi production in ultrarelativistic heavy ion collisions,''
Phys. Lett. B \textbf{523}, 60-66 (2001).
%doi:10.1016/S0370-2693(01)01311-9
%[arXiv:hep-ph/0103124 [hep-ph]].
%215 citations counted in INSPIRE as of 03 Jan 2022
  
%\cite{Peskin:1979va}
\bibitem{Peskin:1979va}
M.~E.~Peskin,
%``Short Distance Analysis for Heavy Quark Systems. 1. Diagrammatics,''
Nucl. Phys. B \textbf{156}, 365-390 (1979).
%doi:10.1016/0550-3213(79)90199-8
%430 citations counted in INSPIRE as of 04 Oct 2021

%\cite{Bhanot:1979vb}
\bibitem{Bhanot:1979vb}
G.~Bhanot and M.~E.~Peskin,
%``Short Distance Analysis for Heavy Quark Systems. 2. Applications,''
Nucl. Phys. B \textbf{156}, 391-416 (1979).
%doi:10.1016/0550-3213(79)90200-1
%397 citations counted in INSPIRE as of 04 Oct 2021

%\cite{Satz:2005hx}
\bibitem{Satz:2005hx}
H.~Satz,
%``Colour deconfinement and quarkonium binding,''
J. Phys. G \textbf{32}, R25 (2006).
%doi:10.1088/0954-3899/32/3/R01
%[arXiv:hep-ph/0512217 [hep-ph]].
%286 citations counted in INSPIRE as of 29 Nov 2021  

%\cite{Zhao:2020jqu}
\bibitem{Zhao:2020jqu}
J.~Zhao, K.~Zhou, S.~Chen and P.~Zhuang,
%``Heavy flavors under extreme conditions in high energy nuclear collisions,''
Prog. Part. Nucl. Phys. \textbf{114}, 103801 (2020).
%doi:10.1016/j.ppnp.2020.103801
%[arXiv:2005.08277 [nucl-th]].
%27 citations counted in INSPIRE as of 29 Oct 2021


%\cite{STAR:2017kkh}
\bibitem{STAR:2017kkh}
L.~Adamczyk \textit{et al.} [STAR],
%``Measurement of $D^0$ Azimuthal Anisotropy at Midrapidity in Au+Au Collisions at $\sqrt{s_{NN}}$=200  GeV,''
Phys. Rev. Lett. \textbf{118}, no.21, 212301 (2017).
%doi:10.1103/PhysRevLett.118.212301
%[arXiv:1701.06060 [nucl-ex]].
%115 citations counted in INSPIRE as of 05 Oct 2021

%\cite{Grandchamp:2003uw}
\bibitem{Grandchamp:2003uw}
L.~Grandchamp, R.~Rapp and G.~E.~Brown,
%``In medium effects on charmonium production in heavy ion collisions,''
Phys. Rev. Lett. \textbf{92}, 212301 (2004).
%doi:10.1103/PhysRevLett.92.212301
%[arXiv:hep-ph/0306077 [hep-ph]].
%200 citations counted in INSPIRE as of 08 Oct 2021

%\cite{Song:2012at}
\bibitem{Song:2012at}
T.~Song, K.~C.~Han and C.~M.~Ko,
%``Charmonium production from nonequilibrium charm and anticharm quarks in quark-gluon plasma,''
Phys. Rev. C \textbf{85}, 054905 (2012).
%doi:10.1103/PhysRevC.85.054905
%[arXiv:1203.2964 [nucl-th]].
%17 citations counted in INSPIRE as of 08 Oct 2021

%\cite{Grandchamp:2002wp}
\bibitem{Grandchamp:2002wp}
L.~Grandchamp and R.~Rapp,
%``Charmonium suppression and regeneration from SPS to RHIC,''
Nucl. Phys. A \textbf{709}, 415-439 (2002).
%doi:10.1016/S0375-9474(02)01027-8
%[arXiv:hep-ph/0205305 [hep-ph]].
%207 citations counted in INSPIRE as of 08 Oct 2021

%\cite{Zhao:2007hh}
\bibitem{Zhao:2007hh}
X.~Zhao and R.~Rapp,
%``Transverse Momentum Spectra of $J/\psi$ in Heavy-Ion Collisions,''
Phys. Lett. B \textbf{664}, 253-257 (2008).
%doi:10.1016/j.physletb.2008.03.068
%[arXiv:0712.2407 [hep-ph]].
%143 citations counted in INSPIRE as of 08 Oct 2021

%\cite{Andronic:2006ky}
\bibitem{Andronic:2006ky}
A.~Andronic, P.~Braun-Munzinger, K.~Redlich and J.~Stachel,
%``Statistical hadronization of heavy quarks in ultra-relativistic nucleus-nucleus collisions,''
Nucl. Phys. A \textbf{789}, 334-356 (2007).
%doi:10.1016/j.nuclphysa.2007.02.013
%[arXiv:nucl-th/0611023 [nucl-th]].
%210 citations counted in INSPIRE as of 08 Oct 2021

%\cite{Kostyuk:2005zd}
\bibitem{Kostyuk:2005zd}
A.~P.~Kostyuk,
%``Double, triple and hidden charm production in the statistical coalescence model,''
arXiv:nucl-th/0502005 [nucl-th].
%7 citations counted in INSPIRE as of 18 Oct 2021

  %\cite{Miller:2007ri}
\bibitem{Miller:2007ri}
M.~L.~Miller, K.~Reygers, S.~J.~Sanders and P.~Steinberg,
%``Glauber modeling in high energy nuclear collisions,''
Ann. Rev. Nucl. Part. Sci. \textbf{57}, 205-243 (2007).
%doi:10.1146/annurev.nucl.57.090506.123020
%[arXiv:nucl-ex/0701025 [nucl-ex]].
%1297 citations counted in INSPIRE as of 08 Jun 2020

%\cite{ALICE:2011aa}
\bibitem{ALICE:2011aa}
B.~Abelev \textit{et al.} [ALICE],
%``Measurement of charm production at central rapidity in proton-proton collisions at $\sqrt{s} = 7$ TeV,''
JHEP \textbf{01}, 128 (2012).
%doi:10.1007/JHEP01(2012)128
%[arXiv:1111.1553 [hep-ex]].
%290 citations counted in INSPIRE as of 08 Jun 2020

%\cite{Cacciari:1998it}
\bibitem{Cacciari:1998it}
M.~Cacciari, M.~Greco and P.~Nason,
%``The P(T) spectrum in heavy flavor hadroproduction,''
JHEP \textbf{05}, 007 (1998).
%doi:10.1088/1126-6708/1998/05/007
%[arXiv:hep-ph/9803400 [hep-ph]].
%773 citations counted in INSPIRE as of 09 Jun 2020

%\cite{Cacciari:2001td}
\bibitem{Cacciari:2001td}
M.~Cacciari, S.~Frixione and P.~Nason,
%``The p(T) spectrum in heavy flavor photoproduction,''
JHEP \textbf{03}, 006 (2001).
%doi:10.1088/1126-6708/2001/03/006
%[arXiv:hep-ph/0102134 [hep-ph]].
%369 citations counted in INSPIRE as of 09 Jun 2020

 %\cite{Zha:2015eca}
\bibitem{Zha:2015eca}
W.~Zha, B.~Huang, R.~Ma, L.~Ruan, Z.~Tang, Z.~Xu, C.~Yang, Q.~Yang and S.~Yang,
%``Systematic study of the experimental measurements on J/\ensuremath{\psi} cross sections and kinematic distributions in p+p collisions at different energies,''
Phys. Rev. C \textbf{93}, no.2, 024919 (2016).
%doi:10.1103/PhysRevC.93.024919
%[arXiv:1506.08985 [hep-ex]].
%11 citations counted in INSPIRE as of 11 Sep 2021

%\cite{STAR:2018smh}
\bibitem{STAR:2018smh}
J.~Adam \textit{et al.} [STAR],
%``$J/\psi$ production cross section and its dependence on charged-particle multiplicity in $p + p$ collisions at $\sqrt{s}$ = 200 GeV,''
Phys. Lett. B \textbf{786}, 87-93 (2018).
%doi:10.1016/j.physletb.2018.09.029
%[arXiv:1805.03745 [hep-ex]].
%21 citations counted in INSPIRE as of 05 Oct 2021

%\cite{ParticleDataGroup:2020ssz}
\bibitem{ParticleDataGroup:2020ssz}
P.~A.~Zyla \textit{et al.} [Particle Data Group],
%``Review of Particle Physics,''
PTEP \textbf{2020}, no.8, 083C01 (2020).
%doi:10.1093/ptep/ptaa104
%2558 citations counted in INSPIRE as of 03 Jan 2022

%\cite{Mueller:1985wy}
\bibitem{Mueller:1985wy}
A.~H.~Mueller and J.~w.~Qiu,
%``Gluon Recombination and Shadowing at Small Values of x,''
Nucl. Phys. B \textbf{268}, 427-452 (1986).
%doi:10.1016/0550-3213(86)90164-1
%1387 citations counted in INSPIRE as of 09 Jun 2020

%\cite{Cronin:1974zm}
\bibitem{Cronin:1974zm}
J.~Cronin, H.~J.~Frisch, M.~Shochet, J.~Boymond, R.~Mermod, P.~Piroue and R.~L.~Sumner,
%``Production of hadrons with large transverse momentum at 200, 300, and 400 GeV,''
Phys. Rev. D \textbf{11}, 3105-3123 (1975).
%doi:10.1103/PhysRevD.11.3105
%1028 citations counted in INSPIRE as of 09 Jun 2020

%\cite{Hufner:1988wz}
\bibitem{Hufner:1988wz}
J.~Hufner, Y.~Kurihara and H.~Pirner,
%``Gluon Multiple Scattering and the Transverse Momentum Dependence of $J/\psi$ Production in Nucleus Nucleus Collisions,''
Acta Phys. Slov. \textbf{39}, 281-287 (1989).
%doi:10.1016/0370-2693(88)91423-2

%\cite{Gerschel:1988wn}
\bibitem{Gerschel:1988wn}
C.~Gerschel and J.~Hufner,
%``A Contribution to the Suppression of the J/psi Meson Produced in High-Energy Nucleus Nucleus Collisions,''
Phys. Lett. B \textbf{207}, 253-256 (1988).
%doi:10.1016/0370-2693(88)90570-9
%181 citations counted in INSPIRE as of 09 Jun 2020

%\cite{Huefner:2002tt}
\bibitem{Huefner:2002tt}
J.~Huefner and P.~f.~Zhuang,
%``Time structure of anomalous J / psi and psi-prime suppression in nuclear collisions,''
Phys. Lett. B \textbf{559}, 193-200 (2003).
%doi:10.1016/S0370-2693(03)00347-2
%[arXiv:nucl-th/0208004 [nucl-th]].
%22 citations counted in INSPIRE as of 04 Oct 2021

%\cite{NA50:2003tdy}
\bibitem{NA50:2003tdy}
N.~S.~Topilskaya \textit{et al.} [NA50],
%``Transverse momentum distribution of J/psi produced in Pb Pb and p A interactions at the CERN SPS,''
Nucl. Phys. A \textbf{715}, 675-678 (2003).
%doi:10.1016/S0375-9474(02)01464-1
%29 citations counted in INSPIRE as of 18 Oct 2021

%\cite{PHENIX:2007tnc}
\bibitem{PHENIX:2007tnc}
A.~Adare \textit{et al.} [PHENIX],
%``Cold Nuclear Matter Effects on J/Psi as Constrained by Deuteron-Gold Measurements at s(NN)**(1/2) = 200-GeV,''
Phys. Rev. C \textbf{77}, 024912 (2008)
[erratum: Phys. Rev. C \textbf{79}, 059901 (2009)].
%doi:10.1103/PhysRevC.77.024912
%[arXiv:0903.4845 [nucl-ex]].
%161 citations counted in INSPIRE as of 18 Oct 2021

%\cite{Gavin:1996bx}
\bibitem{Gavin:1996bx}
S.~Gavin, P.~L.~McGaughey, P.~V.~Ruuskanen and R.~Vogt,
%``How to find charm in nuclear collisions at RHIC and CERN LHC,''
Phys. Rev. C \textbf{54}, 2606-2623 (1996).
%doi:10.1103/PhysRevC.54.2606
%[arXiv:hep-ph/9604369 [hep-ph]].
%92 citations counted in INSPIRE as of 03 Jan 2022

%\cite{Vogt:2010aa}
\bibitem{Vogt:2010aa}
R.~Vogt,
%``Cold Nuclear Matter Effects on $J/\psi$ and $\Upsilon$ Production at the LHC,''
Phys. Rev. C \textbf{81}, 044903 (2010).
%doi:10.1103/PhysRevC.81.044903
%[arXiv:1003.3497 [hep-ph]].
%161 citations counted in INSPIRE as of 03 Jan 2022

%\cite{Norton:2003cb}
\bibitem{Norton:2003cb}
P.~Norton,
%``The EMC effect,''
Rept. Prog. Phys. \textbf{66}, 1253-1297 (2003).
%doi:10.1088/0034-4885/66/8/201
%124 citations counted in INSPIRE as of 16 Jun 2020

%\cite{Teaney:2003kp}
\bibitem{Teaney:2003kp}
D.~Teaney,
%``The Effects of viscosity on spectra, elliptic flow, and HBT radii,''
Phys. Rev. C \textbf{68}, 034913 (2003).
%doi:10.1103/PhysRevC.68.034913
%[arXiv:nucl-th/0301099 [nucl-th]].
%759 citations counted in INSPIRE as of 18 Oct 2021

%\cite{Lacey:2006bc}
\bibitem{Lacey:2006bc}
R.~A.~Lacey, N.~N.~Ajitanand, J.~M.~Alexander, P.~Chung, W.~G.~Holzmann, M.~Issah, A.~Taranenko, P.~Danielewicz and H.~Stoecker,
%``Has the QCD Critical Point been Signaled by Observations at RHIC?,''
Phys. Rev. Lett. \textbf{98}, 092301 (2007).
%doi:10.1103/PhysRevLett.98.092301
%[arXiv:nucl-ex/0609025 [nucl-ex]].
%264 citations counted in INSPIRE as of 18 Oct 2021

%\cite{Monnai:2019hkn}
\bibitem{Monnai:2019hkn}
A.~Monnai, B.~Schenke and C.~Shen,
%``Equation of state at finite densities for QCD matter in nuclear collisions,''
Phys. Rev. C \textbf{100}, no.2, 024907 (2019).
%doi:10.1103/PhysRevC.100.024907
%[arXiv:1902.05095 [nucl-th]].
%48 citations counted in INSPIRE as of 27 Oct 2021

%\cite{Monnai:2021kgu}
\bibitem{Monnai:2021kgu}
A.~Monnai, B.~Schenke and C.~Shen,
%``QCD Equation of State at Finite Chemical Potentials for Relativistic Nuclear Collisions,''
Int. J. Mod. Phys. A \textbf{36}, no.07, 2130007 (2021).
%doi:10.1142/S0217751X21300076
%[arXiv:2101.11591 [nucl-th]].
%13 citations counted in INSPIRE as of 27 Oct 2021

%\cite{Denicol:2018wdp}
\bibitem{Denicol:2018wdp}
G.~S.~Denicol, C.~Gale, S.~Jeon, A.~Monnai, B.~Schenke and C.~Shen,
%``Net baryon diffusion in fluid dynamic simulations of relativistic heavy-ion collisions,''
Phys. Rev. C \textbf{98}, no.3, 034916 (2018).
%doi:10.1103/PhysRevC.98.034916
%[arXiv:1804.10557 [nucl-th]].
%87 citations counted in INSPIRE as of 27 Oct 2021

%\cite{Kharzeev:2000ph}
\bibitem{Kharzeev:2000ph}
D.~Kharzeev and M.~Nardi,
%``Hadron production in nuclear collisions at RHIC and high density QCD,''
Phys. Lett. B \textbf{507}, 121-128 (2001).
%doi:10.1016/S0370-2693(01)00457-9
%[arXiv:nucl-th/0012025 [nucl-th]].
%798 citations counted in INSPIRE as of 21 Sep 2021

%\cite{Wu:2021fjf}
\bibitem{Wu:2021fjf}
X.~Y.~Wu, G.~Y.~Qin, L.~G.~Pang and X.~N.~Wang,
%``(3+1)-D viscous hydrodynamics CLVisc at finite net baryon density: identified particle spectra, anisotropic flows and flow fluctuations across BES energies,''
[arXiv:2107.04949 [hep-ph]].
%2 citations counted in INSPIRE as of 27 Oct 2021

%\cite{ALICE:2012fjm}
\bibitem{ALICE:2012fjm}
B.~Abelev \textit{et al.} [ALICE],
%``Measurement of inelastic, single- and double-diffraction cross sections in proton--proton collisions at the LHC with ALICE,''
Eur. Phys. J. C \textbf{73}, no.6, 2456 (2013).
%doi:10.1140/epjc/s10052-013-2456-0
%[arXiv:1208.4968 [hep-ex]].
%301 citations counted in INSPIRE as of 18 Oct 2021

%\cite{Sjostrand:2006za}
\bibitem{Sjostrand:2006za}
T.~Sjostrand, S.~Mrenna and P.~Z.~Skands,
%``PYTHIA 6.4 Physics and Manual,''
JHEP \textbf{05}, 026 (2006).
%doi:10.1088/1126-6708/2006/05/026
%[arXiv:hep-ph/0603175 [hep-ph]].
%12035 citations counted in INSPIRE as of 18 Oct 2021

 %\cite{Shen:2017bsr}
\bibitem{Shen:2017bsr} 
  C.~Shen and B.~Schenke,
  %``Dynamical initial state model for relativistic heavy-ion collisions,''
  Phys.\ Rev.\ C {\bf 97}, no. 2, 024907 (2018).
      
\bibitem{Shen:2017fnn} 
  C.~Shen and B.~Schenke,
  %``Initial state and hydrodynamic modeling of heavy-ion collisions at RHIC BES energies,''
  PoS CPOD {\bf 2017}, 006 (2018).
 
 %\cite{Schenke:2010nt}
\bibitem{Schenke:2010nt}
B.~Schenke, S.~Jeon and C.~Gale,
%``(3+1)D hydrodynamic simulation of relativistic heavy-ion collisions,''
Phys. Rev. C \textbf{82}, 014903 (2010).
%doi:10.1103/PhysRevC.82.014903
%[arXiv:1004.1408 [hep-ph]].
%328 citations counted in INSPIRE as of 27 Oct 2021

%\cite{Andronic:2008gu}
\bibitem{Andronic:2008gu}
A.~Andronic, P.~Braun-Munzinger and J.~Stachel,
%``Thermal hadron production in relativistic nuclear collisions: The Hadron mass spectrum, the horn, and the QCD phase transition,''
Phys. Lett. B \textbf{673}, 142-145 (2009)
[erratum: Phys. Lett. B \textbf{678}, 516 (2009)].
%doi:10.1016/j.physletb.2009.06.021
%[arXiv:0812.1186 [nucl-th]].
%475 citations counted in INSPIRE as of 03 Jan 2022

%\cite{Cleymans:2005xv}
\bibitem{Cleymans:2005xv}
J.~Cleymans, H.~Oeschler, K.~Redlich and S.~Wheaton,
%``Comparison of chemical freeze-out criteria in heavy-ion collisions,''
Phys. Rev. C \textbf{73}, 034905 (2006).
%doi:10.1103/PhysRevC.73.034905
%[arXiv:hep-ph/0511094 [hep-ph]].
%619 citations counted in INSPIRE as of 04 Jan 2022

%\cite{STAR:2012wnc}
\bibitem{STAR:2012wnc}
L.~Adamczyk \textit{et al.} [STAR],
%``$J/\psi$ production at high transverse momenta in $p+p$ and Au+Au collisions at $\sqrt{s_{NN}} = 200$ GeV,''
Phys. Lett. B \textbf{722}, 55-62 (2013).
%doi:10.1016/j.physletb.2013.04.010
%[arXiv:1208.2736 [nucl-ex]].
%121 citations counted in INSPIRE as of 18 Oct 2021



  
\end{thebibliography}
\end{document}